# Modular Hardware Design with Timeline Types

RACHIT NIGAM, Cornell University, USA

PEDRO HENRIQUE AZEVEDO DE AMORIM, Cornell University, USA

ADRIAN SAMPSON, Cornell University, USA

Modular design is a key challenge for enabling large-scale reuse of hardware modules. Unlike software, however, hardware designs correspond to physical circuits and inherit constraints from them. Timing constraints—which cycle a signal arrives, when an input is read—and structural constraints—how often a multiplier accepts new inputs—are fundamental to hardware interfaces. Existing hardware design languages do not provide a way to encode these constraints; a user must read documentation, build scripts, or in the worst case, a module's implementation to understand how to use it. We present Filament, a language for modular hardware design that supports the specification and enforcement of timing and structural constraints for statically scheduled pipelines. Filament uses *timeline types*, which describe the intervals of clock-cycle time when a given signal is available or required. Filament enables *safe composition* of hardware modules, ensures that the resulting designs are correctly pipelined, and predictably lowers them to efficient hardware.

## 1 INTRODUCTION

Like software languages, interfaces in hardware description languages (HDLs) simply consist of arguments and simple datatypes. Unlike software, hardware inherits constraints from the underlying physical circuits—the inputs are used and outputs are available during specific cycles, and new inputs may only be provided when the circuit can process them. The rudimentary interfaces of existing HDLs fail to capture these constraints, making modular design difficult. To approach the reusability of software library ecosystems, HDLs need a systematic way to encode these requirements for hardware modules.

Modern languages for hardware design fall into three categories. Embedded HDLs (eHDLs) use software host languages for metaprogramming [4, 16, 29, 34, 41]. Accelerator design languages (ADLs) [22, 25, 26, 31, 38, 51] are higher-level languages that expose new abstractions and compile to HDLs. Finally, traditional HDLs, such as SystemVerilog and VHDL, are the de facto standard for hardware design and interacting with hardware toolchains. ADLs and eHDLs must be compiled to HDLs to interact with hardware design toolchains and integrate proprietary black-box implementations of optimized primitives. Furthermore, ADLs are often limited to a single domain, such as image processing, and can therefore benefit from defining a foreign function interface to interact with eHDLs and other ADLs.

Composition in HDLs is challenging because interfaces only expose the names and value types of input–output ports. However, efficient integration requires knowledge of timing behavior: the number of cycles required to produce and consume outputs and inputs respectively, and whether a module can be pipelined. In current HDLs, this timing information is latent. It appears in verbose documentation files, encoded as constraints in build scripts [47], or nowhere at all, requiring users to read the implementation of each module to understand how to use it. An alternative is to rely exclusively on latency-insensitive interfaces which eliminate all timing sensitivity. Instead, the producer signals when its output is valid and the consumer signals when it is ready to accept new inputs. While latency-insensitive interfaces are flexible, they are also inefficient [37], especially for *statically scheduled* modules which always take the same number of clock cycles to produce outputs and accept new inputs.

The key to an ecosystem of reusable hardware is a low-level mechanism to *safely* and *efficiently* compose hardware modules. *Safe composition* requires modules to specify and check timing details such as latency and pipelinability, while *efficiency* requires that the interfaces do not add substantial overheads. The efficiency requirement also rules out wrapping statically scheduled modules

```
module Add(a: 32, b: 32) -> (o: 32);
module Mul(a: 32, b: 32) -> (o: 32);
module Mux(sel: 1, a: 32, b: 32) -> (o: 32);
module ALU(op: 1, l: 32, r: 32) -> (o: 32) {
  Mul M(l, r); Add A(l, r);
  Mux Mx(op, A.out, M.out); o = Mx.out; }
```

(a) HDL implementation of ALU

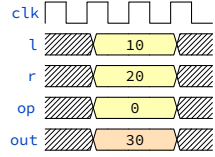

(b) Addition

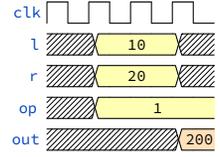

(c) Multiplication

Fig. 1. ALU implementation and waveforms generated when executing addition and multiplication.

with a latency-insensitive interface; instead, we would like to use the clock signal to synchronize usage of the modules. The effect is a clear way to integrate hardware, regardless of whether it was written in an eHDL, generated by an ADL, or implemented as a proprietary black box module.

Our solution is *timeline types*, which compactly encode latency and throughput properties of statically scheduled hardware pipelines. Static pipelines have data-independent timing behavior and encompass a large class of efficient hardware designs [22, 25, 26, 30] including the pipelines generated by most high-level synthesis (HLS) tools [11, 38, 42, 51]. Our type system, inspired by separation logic [46], proves that pipelined execution of a module is safe by ensuring that all timing constraints are satisfied. Our contributions are as follows:

- We provide a characterization of pipelining constraints for static pipelines and model them using *timeline types* in an HDL called Filament.
- We formalize these pipelining constraints using a *log-based semantics* of hardware and prove our type system is sound with respect to the model.
- We demonstrate that Filament can integrate designs from several hardware generators [22, 30, 49] using timeline types.
- We show that Filament designs use fewer resources and run at faster frequencies than those generated by hardware generators.

## 2 EXAMPLE

We will discuss the challenges associated with compositional hardware design by implementing a pipelined arithmetic logic unit (ALU).

### 2.1 Traditional Hardware Description Languages

Figure 1a shows the implementation of the ALU in a traditional HDL. The interfaces for the modules specify the inputs and outputs along with their bitwidths. The ALU's circuit consists of an adder and a multiplier, which perform their computations in parallel, and a multiplexer, which selects between the two outputs using the op signal.

We will use waveform diagrams to understand the execution behavior of this module. A waveform diagram explains the flow of signals in the circuit over time and usually with respect to the global clock signal. Figure 1b shows the waveform generated when the ALU is provided with the inputs 10 and 20 and the op code 0. Note that the output 30 is produced in the same cycle as the inputs. However, Figure 1c shows what happens when we attempt to execute the multiplication operation by setting op to 1. The timing behavior of the ALU changes—the product is produced two cycles after the input is provided. Additionally, if the op is not asserted for an additional cycle, the output is wrong. The problem is that an adder is *combinational*—it produces its output in the same cycle as the inputs—while a multiplier is *sequential*—it takes several cycles to produce its output. op is required for an extra cycle because the multiplier output is produced later than the adder and the multiplexer needs to select the correct output in a later cycle using the op input.

The interfaces for ALU, the adder, and the multiplier do not capture these details. One option to sidestep this problem is to "wrap" every module in a *latency-insensitive* interface, such as ready–valid handshaking. But these interfaces incur overhead that can be prohibitive for fine-grained composition [37]. This paper aims to specify efficient, *latency-sensitive* interfaces based on clock cycles and to statically rule out misuses of these interfaces.

## 2.2  Filament

Filament is an HDL that allows users to directly *specify* and *check* the timing behavior of their modules. Each component can be parameterized by multiple events which are used to specify its timing behavior. Our ALU implementation has behaves unpredictably because adders and multipliers have different timing behavior. Filament allows us to encode their timing behavior explicitly using *events* which parameterize modules:

```
extern comp Add<T>(
  @interface[T] go: 1, @[T, T+1] left: 32, @[T, T+1] right: 32) -> (@[T, T+1] out: 32);
extern comp Mult<T>(
  @interface[T] go: 1, @[T, T+1] left: 32, @[T, T+1] right: 32) -> (@[T+2, T+3] out: 32);
```

Both components use the event *T* to specify their timing behavior. The adder is *combinational*—it produces outputs in the same cycle as the inputs. This fact is encoded by the *availability intervals* of the inputs and outputs: the inputs are provided in the half-open interval $[T, T + 1)$, which corresponds to the first cycle of execution of the component, and the output is produced during the same interval. In contrast, a multiplier is *sequential*—it takes two cycles to produce its output. This is encoded by stating that the output is available in the interval $[T + 2, T + 3)$, two cycles after the inputs are provided in the interval $[T, T + 1)$. In order to signal that the event *T* has occurred, a user of these modules must set the *interface port* go to 1, provide the inputs according to their required intervals, and read the output when they are available. Multiplexers (not shown) are also combinational and take all their inputs in the same cycle.

Like our HDL implementation, our Filament implementation of the ALU explicitly instantiates all the hardware resources it needs to use. The key difference is how Filament expresses the use of the hardware instances through *invocations*. An invocation schedules the execution of a hardware instance using a particular set of events and provides all inputs. For example, the invocation a0 of the adder A is scheduled using the event *G*. By naming uses, Filament can check the timing behavior of the module. There is no assignment for the go port of the adder—it is automatically inserted by the compiler using the scheduling event *G*. Invocations are a logical construct that are compiled away by Filament (Section 5). Similarly, the multiplier and multiplexer are also scheduled using the event *G*. Instead of using outputs from the instance, the

```
comp ALU<G>(
  @interface[G] en: 1
  @[G, G+1] op: 1,
  @[G, G+1] l: 32,
  @[G, G+1] r: 32,
) -> (@[G+2, G+3] o: 32) {
  A := new Add; M := new Mult;
  Mx := new Mux;
  a0 := A<G>(l, r);
  m0 := M<G>(l, r);
  mux := Mux<G>(
    op, m0.out, a0.out);
  o = mux.out; }
```

multiplexer uses the ports on the invocations, reflecting the output from a particular use.

## 2.3  Checking Timing Behavior

However, when we attempt to compile this program, Filament gives us the following error:

```
mux := Mux<G>(op, m0.out, a0.out);
Available for [G+2, G+3] but required during [G, G+1]
```

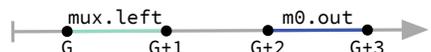

The error states that the use of our multiplexer expects all of its inputs during $[G, G+1)$ while the multiplier's output, `m0.out`, is available in $[G+2, G+3)$. Filament requires that all inputs be available for at least as long as the corresponding argument's requirement. This was the problem in our original HDL design (Section 2.1)—the output of the adder is available in a different cycle from the multiplier which results in unexpected timing behavior. Filament's type system statically catches this error.

The solution is to use *registers* to store values and make them available in future cycles. A register's signature captures its timing behavior—the output is available one cycle after the input[1]:

```
comp Reg<G>(@interface[G] en: 1, @[G, G+1] in: 32) -> (@[G+1, G+2] out: 32)
```

The corrected implementation uses two registers to make the sum available in the same cycle as the multiplier. The outputs from the first and second registers are available in $[G+1, G+2)$ and $[G+2, G+3)$, respectively. We schedule the execution of the multiplexer in cycle $G+2$ when both the outputs are available. This design is still problematic because the `op` is only available in $[G, G+1)$ while the multiplexer reads it in $[G+2, G+3)$. We fix this by making `op` signal available in $[G, G+3)$. This results in a correct ALU implementation.

However, it is not clear when the ALU is ready to accept new inputs: should we wait till outputs are produced or can the module process multiple inputs in parallel?

```
comp ALU<G>(@[G, G+3] op: 32, ...) {
  a0 := A<G>(l, r);  R0 := new Reg;  R1 := new Reg;
  r0 := R0<G>(a0.out);  r1 := R1<G+1>(r0.out);
  mux := Mux<G+2>(op, r1.out, m0.out); ... }
```

## 2.4  Pipelining

*Pipelining* is a common optimization that enables hardware to process multiple inputs in parallel. A sequential module processes its inputs one at a time (Diagram (a)), while pipelined module can overlap the processing of multiple inputs (Diagram (b)).

Pipelining is challenging because it requires reasoning about the interaction between multiple, concurrent executions of the same physical resources—correctly pipelining requires using values from the correct pipeline stage and ensuring there are no *structural hazards*, i.e., there are no conflicting uses of internal components.

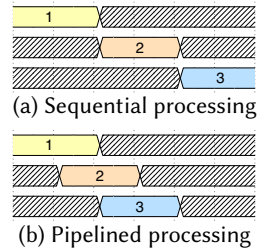

(a) Sequential processing

(b) Pipelined processing

Filament presents a concise solution: each event has an associated *delay* that specifies how many cycles to wait before accepting new inputs. We can update the signature of the adder and multiplier to reflect this. Since the adder is combinational, it can accept new inputs every cycle. However, the multiplier accepts new inputs every 3 cycles. For user-level components, Filament ensures that the *delay* for each event is correct, i.e., the component can be correctly pipelined. We'll redesign our ALU to be pipelined and accept new inputs every cycle by specifying that the delay of $G$ is 1. Since we know our design is not pipelined, Filament will generate errors explaining why the design cannot be pipelined.

```
comp Add<T:1>(...)
comp Mult<T:3>(...)
comp ALU<G:1>(...)
```

```
comp ALU<G:1>(
        Event may retrigger every cycle
  @[G, G+3] op: 1, Signal lasts for 3 cycles
```

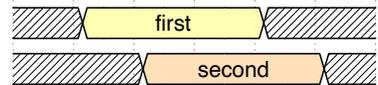

---

[1] This is a simplified interface for a register. Full interface provided in Section 3.6.

```
def init(left) -> (acc, q):
    # Initialize the computation
def nxt(a, q, div) -> (an, qn):
    # One step of the computation
def div(l, r):
    (qn, an) = init(l)
    for _ in range(0, 8):
        (qn, an) = nxt(an, qn, r)
    return qn
```

(a) Pseudocode for restoring division.

```
comp Comb<G: 1>(...) -> (
    @[G, G+1] out: 8) {
i := new Init<G>(left);
n0 := new Nxt<G>(i.A, i.Q, r);
...
n7 := new Nxt<G>(n6.A, n6.Q, r);
out = n7.Q;
```

(b) Fully combinational divider.

```
comp Pipe<G: 1>(...) -> (@[G+7, G+8] q: 8) {
i := new Init<G>(left); // Instantiate and invoke
n0 := new Nxt<G>(i.A, i.Q, r);
ra0 := new Reg<G>(n0.A);
rq0 := new Reg<G>(n0.Q);
n2 := new Nxt<G+1>(ra0.out, div, rq0.out);
...
out = n7.Q;
```

(c) Pipelined divider. Instances scheduled in successive cycles.

```
comp Iter<G: 8>(...) -> (@[G+7, G+8] q: 8) {
I := new Init<G>(left);
N := new Nxt; RA := new Reg; RQ := new Reg;
n0 := new N<G>(i.A, i.Q, r);
ra0 := new RA<G>(n0.A); rq0 := new RQ<G>(n0.Q);
n1 := new N<G+1>(ra0.out, rq0.out);
ra1 := new RA<G+1>(n1.A); rq1 := new RQ<G+1>(n1.Q); ...
out = n7.Q;
```

(d) Iterative divider. Components reused over multiple cycles.

Fig. 2. Implementations of 8-bit restoring division demonstrating area-throughput trade-off. Filament's type system ensures that each implementation is correctly pipelined and introduces no resource reuse conflicts.

Our first problem is that the signature requires input signal op to be available for three cycles whereas the pipeline may trigger every cycle. The waveform diagram demonstrates the problem—the input for op from the first iteration will overlap with the input for the second iteration. However, op is a *physical port* in a circuit and can only hold one value at a time; this is a fundamental physical constraint of hardware design. Filament requires that the delay of an event is at least as long as the length of any availability interval that uses it; we must make op's availability interval 1-cycle long. We choose $[G + 2, G + 3]$ since the multiplexer uses op during this interval.

Next, Filament complains that while our ALU pipeline may accept new inputs every cycle, the multiplier M can accept new inputs every 3 cycles. This is a fundamental limitation of the multiplier circuit we're using; to fix it, we must use a different multiplier. Filament catches yet another pipelining bug that arises from composition: every subcomponent used in a pipeline must be able to process inputs at least as often as the pipeline itself. Fixing this will result in a correct, fully pipelined ALU. A key goal of Filament is to ensure that changing the pipelining behavior of a component does not create additional bugs—the pipelined ALU, like the sequential ALU, only uses signals when they are semantically valid.

```
comp Mult<T: 3>(
Event may retrigger every 3 cycles
comp ALU<G: 1>(
Event may retrigger every cycle
    m0 := M<G>(l, r);
Cannot safely pipeline
```

```
comp ALU<G: 1>(@interface[G] en: 1, @[G+2, G+3] op: 1, ...) {
A := new Add; Mx := new Mux; R0 := new Reg; R1 := new Reg; FM := new FastMult; // delay = 1
a0 := A<G>(l, r); r0 := R0<G>(a0.out); r1 := R1<G+1>(r0.out); m0 := FM<G>(l, r);
mux := Mux<G>(op, r1.out, m0.out); o = mux.out; }
```

## 2.5 Area-Throughput Trade-offs with Filament

While pipelining improves the throughput of a component, it also increases its resource usage. For large circuits, like floating-point multipliers, it often makes sense to reuse the same circuit over multiple clock cycles. However, circuit reuse affects pipelining behavior: the ability of a component to start new iterations depends upon how sub-components are being shared. Filament's type system tracks resource reuse and ensures that a well-typed component does not create structural hazards for reuses components.

To demonstrate how Filament enables safe exploration of *area-throughput trade-offs*, we implement three different versions of a divider using a restoring division algorithm (Figure 2a). The combinational components `Init` and `Nxt` compute a quotient (`.Q`) and an accumulator value (`.A`). For an 8-bit value, we must apply `Nxt` 8 times.

***Combinational divider***. Figure 2b implements a combinational divider which computes the output in the same cycle when the inputs are provided. All `Nxt` instances are scheduled using the event *G* which means that they'll execute in the same cycle. While the latency of the design is 1, it is quite inefficient because it schedules a lot of complex logic in the same clock cycle and forces the design to operate at a low frequency. However, combinational designs are a good starting point to ensure that our algorithm is correct.

***Pipelined divider***. To make our design run at a higher frequency, we can pipeline it by scheduling each `Nxt` instance to execute in successive cycles. To correctly forward the values, we instantiate registers to hold onto values of the quotient and the accumulator for each `Nxt` component. Figure 2c shows the implementation: the delay of the module remains 1, allowing it to process a new value every cycle, but the latency is now 8 cycles unlike the combinational implementation. The pipelining also breaks up the long combinational path allowing the design to operate at a higher frequency.

***Iterative divider***. Both the combinational and pipelined inputs can process a new input every cycle but require a large amount of hardware since they instantiate 8 instances of the `Nxt` component and 16 registers for the pipelined version. We can instead use the same `Nxt` component and registers by implementing an *iterative design*.

We start with our combinational design and change all the invocations to use the same instance *N*. Filament tells us that this design is buggy. We're attempting to send two different inputs into the `Nxt`

```
comp Nxt<T:1>(...)
Delay requires uses to be 1 cycle apart
s0 := N<G>(i.A, div, i.Q); First use
s1 := N<G>(s0.AN, div, s0.QN); Second use
```

instance in the same cycle. However, `Nxt` is a physical circuit and can only process one input every cycle. Therefore, we must schedule the uses of the instance in different cycles and add registers to hold onto the values, similar to the pipelined implementation.

With these changes, Filament complains with a new error message. Since we're sharing the instance `Nxt` over 8 cycles, the divider cannot start processing new inputs every cycle. Again, this is because `Nxt` is a physical circuit that can only process one input a cycle. To fix this, we can change the delay to 8 cycles which guarantees to Filament that the instance will only be run every 8 cycles, resulting in the final design

```
comp Iter<G:1>(...)
Event may trigger every cycle
causing shared uses to conflict
s0 := N<G>(i.A, div, i.Q);
     First use
s7 := N<G+7>(s6.AN, div, s6.QN);
     Last use
```

(Figure 2d). This ensures that all iterations using the instance *N* complete before new inputs are provided. Implicitly, Filament showed us that reusing the instance is a trade-off: while we use fewer resources, our throughput is also reduced since our iterative implementation can only process a new input every 8 cycles compared to every cycle for the pipelined implementation.

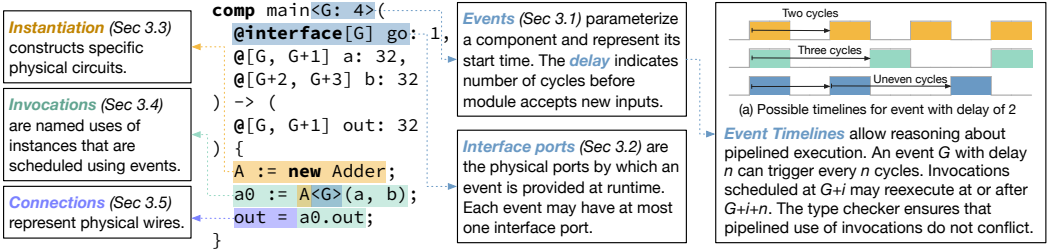

```
comp main<G: 4>(
    @interface[G] go: 1,
    @[G, G+1] a: 32,
    @[G+2, G+3] b: 32
) -> (
    @[G, G+1] out: 32
) {
    A := new Adder;
    a0 := A<G>(a, b);
    out = a0.out;
}
```

**Instantiation** (Sec 3.3) constructs specific physical circuits.

**Invocations** (Sec 3.4) are named uses of instances that are scheduled using events.

**Connections** (Sec 3.5) represent physical wires.

**Events** (Sec 3.1) parameterize a component and represent its start time. The *delay* indicates number of cycles before module accepts new inputs.

**Interface ports** (Sec 3.2) are the physical ports by which an event is provided at runtime. Each event may have at most one interface port.

Two cycles

Three cycles

Uneven cycles

(a) Possible timelines for event with delay of 2

**Event Timelines** allow reasoning about pipelined execution. An event $G$ with delay $n$ can trigger every $n$ cycles. Invocations scheduled at $G+i$ may reexecute at or after $G+i+n$. The type checker ensures that pipelined use of invocations do not conflict.

Fig. 3. Overview of the Filament language. Programs are a sequence of *component* definitions which correspond to individual modules. The signature of the component is parameterized using *events*. The body of component consists of three types of statements: Instantiations, connections, and invocations.

## 2.6 Summary

Filament is an HDL for safe design and composition of static pipelines. Specifically, Filament programs can *specify* and *check* timing properties of hardware modules and ensure that:

(1) Values on ports and wires are only read when they are *semantically* valid.
(2) Hardware instances are not used in a conflicting manner.

These properties ensure that the resulting pipelines are safe, i.e., there are no resource conflicts, and efficient, i.e., they can overlap computation as specified by their interface without any overhead. Filament's utility extends to components defined outside the language as well. By giving external modules a type signature, users can safely compose modules. Section 3 overviews the constructs in Filament, Section 4 explains how Filament's type system checks pipeline safety, and Section 5 shows how Filament's high-level constructs are compiled to efficient hardware.

## 3 THE FILAMENT LANGUAGE

Figure 3 gives an overview of the Filament language. Filament's level of abstraction is comparable to *structural* HDLs where computation must be explicitly mapped onto hardware. Filament only has four constructs: components, instantiation, connections, and invocations. The first three have direct analogues in traditional HDLs while invocations are a novel construct.

### 3.1 Events and Timelines

Events are the core abstraction of time in Filament. Instead of using a clock signal, designs use events to schedule computation. The Filament compiler generates efficient, pipelined finite state machines to reify events (Section 5.2).

**Defining events.** There are only two ways to define events: (1) component signatures bind *event variables* like $G$, and (2) users can write *event expressions* such as $G+n$ where $n$ is a constant. Events have a direct relationship to clock: if $G$ occurs at clock cycle $i$, then $G + n$ occurs at clock cycle $i+n$.[2] This relationship with clock is crucial since it allows Filament to represent timing properties of components defined in clock-based HDLs. Adding event variables ($G_0 + G_1$) is disallowed since events correspond to particular clock cycles, and it is meaningless to add them together.

**Timeline interpretation of events.** In order to capture potential resource conflicts from pipelined execution, Filament interprets events as a set of possible timelines. A timeline for an event $G$ with a delay $n$ is any infinite sequence of 1 cycle long clock pulses such that each pulse is at least $n$ cycles apart. Figure 2a shows a set of valid timelines for an event with delay 2. By imbuing events with a

---

[2]All event variables operate in the same clock domain, but this limitation can be removed in the future.

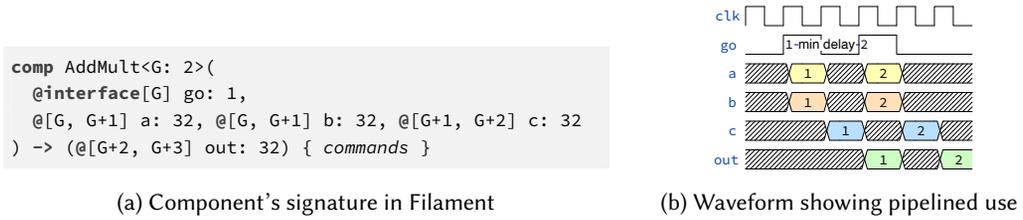

```
comp AddMult<G: 2>(
  @interface[G] go: 1,
  @[G, G+1] a: 32, @[G, G+1] b: 32, @[G+1, G+2] c: 32
) -> (@[G+2, G+3] out: 32) { commands }
```

(a) Component's signature in Filament

(b) Waveform showing pipelined use

Fig. 4. Signature and waveform diagram. The component allows pipelined execution or reuse after two cycles allowing overlapped execution. Shaded regions represent unknown values.

timeline interpretation, Filament can reason about *repeated execution* and consider how pipelined executions may affect each other. By reasoning about such properties, we can define and enforce safety properties for pipelined execution of hardware. Furthermore, the timeline interpretation has a direct relationship to hardware: the delay of an event represents how many cycles a user must wait before providing a new set of inputs. This is usually referred to as the *initiation interval* of a pipeline by hardware designers (Section 4.3).

## 3.2 Components

Filament programs are organized in terms of *components* which describe timing behavior in their signatures and their circuit using a set of commands. Figure 4 shows the signature of a component in Filament (Figure 4a) and a waveform diagram visualizing two sets of inputs being processed in parallel (Figure 4b). The component is parameterized using the event $G$ with a delay of 2 which means that pipelined use can begin two cycles after the previous use.

***Interface ports.*** Hardware components typically have *control ports* which signal when values on *data ports* are valid and that the computation should be performed. Values on control ports are always considered semantically valid while values on data ports are only valid when the corresponding control port is high. Filament distinguishes control ports by defining them as *interface ports*. Interface ports are 1-bit ports that are associated with a particular event. When an interface port is set to 1, it signals to the component that the corresponding event has occurred. For example, setting go to 1 on an AddMult instance (Figure 4a) makes the module start processing the inputs. The availability intervals of all ports that use an event are relative to when the corresponding event's interface port is set to 1. If an event does not have an interface port, then the module can assume that the event triggers every $n$ cycles where $n$ is the event's delay.

***Availability intervals.*** The input and output ports of the component describe their availability in terms of the events bound by a component. For AddMult (Figure 4a), all ports use the event $G$. Availability intervals are *half-open*: for example, the input port a is available during $[G, G + 1)$ which means it available during the first cycle when the component is invoked. Inside the body of a component, an input's availability interval represents a *guarantee* while an output's availability requires a *requirement* that the body must fulfill. When using a component, this is reversed: inputs have requirements that must be fulfilled by the user while outputs have guarantees.

## 3.3 Instances

All computations in a hardware design must be explicitly mapped onto physical circuits. Filament's new keyword allows instantiation of subcomponents. The following program instantiates two instances of the Add component named A0 and A1 that can be used independently. The instantiations do not provide bindings for the `Add`'s event $T$; *invocations* are responsible for providing those and scheduling the execution of an instance.

```
comp Add<T: 1>(@[T, T+1] left: 32, @[T, T+1] right) -> (@[T, T+1] o: 32);
comp AddTwo<G: 1>(...) { A0 := new Add; A1 := new Add; ... }
```

### 3.4 Invocations

Resource reuse in hardware designs is *time-multiplexed*, i.e., different uses of the same resources are scheduled to occur at different times. This is done by building a finite state machine (FSM) using a register and using the output of the register to select which inputs to use. The example program computes $(l \times r)^2$ using a single multiplier using the FSM `F` to forward the inputs $l$ and $r$ into the multiplier

```
F := new Reg; // FSM
F.in = F.out == 0 ? 1 : 0;
M := new Mult; A := new Add;
M.right = F.out == 0 ? r : M.out
M.left = F.out == 1 ? l : M.out
```

in the first cycle and the output of the multiplier in the second cycle. However, the assignment to `M.left` incorrectly forwards the value from `M.out` in the first cycle. Mistakes in the control logic for the FSM do not lead to any visible errors; this error will lead to the data getting silently corrupted and propagating into other parts of the system.

In contrast, every use of an instance in Filament must be explicitly named and scheduled through an invocation. The first invocation of the multiplier `M` is scheduled using the event $G$, uses the inputs $l$ and $r$, and is named `m0`. The second invocation, scheduled one cycle later at $G + 1$, can then use `m0.out` to refer to the output of the first execution and pass it into the multiplier as an input. Because the second invocation is scheduled one cycle later, the input ports have a different requirement: the inputs must be available in the interval $[G + 1, G + 2)$

```
comp Square<T:1>(
  @[T, T+1] left: 32,
  @[T, T+1] right: 32
) -> (
  @[T+1, T+2] out: 32);
M := new Mult;
m0 := M<G>(l, r)
m1 := M<G+1>(
  m0.out, m0.out)
```

as opposed to $[G, G + 1)$ in the first invocation. This allows Filament to check that `m0.out` is semantically valid when it is used as an input to `m1` and that the two uses of the multiplier are scheduled to occur at different times, allowing the compiler to generate correct FSMs to schedule instance reuse. Each invocation only provides inputs for the data ports and elides inputs for the interface ports. During compilation, Filament's compiler automatically infers assignments for the input ports and generates efficient, pipelined FSMs to schedule the invocations (Section 5).

### 3.5 Connections

Filament programs allow ports to be connected and requires that the source is semantically valid for at least as long as the destination. Connections are physically implemented as continuously active wires connecting two ports in the circuit.

```
comp Add<G:1>(@[G, G+3] source: 32) -> (@[G, G+1] dest: 32) { dest = source; }
```

### 3.6 Interfacing with External Components

Filament's `extern` keyword allows the user to provide type-safe wrappers for black box modules by specifying a type signature without a body. Filament's standard library, which provides signatures for components like multipliers and registers, is defined using `extern` components.

***Phantom events***. Phantom events allow Filament to model the behavior of components like adders which are *continuously* active and do not take an explicit enable signal. In the following signature, the event $G$ is a phantom event because there is no corresponding interface port for it in the signature. Section 5.4 describes how user-level components can use phantom events.

```
extern comp Add<G: 1>(@[G, G+1] l: 32, @[G, G+1] r: 32) -> (@[G, G+1] o: 32))
```

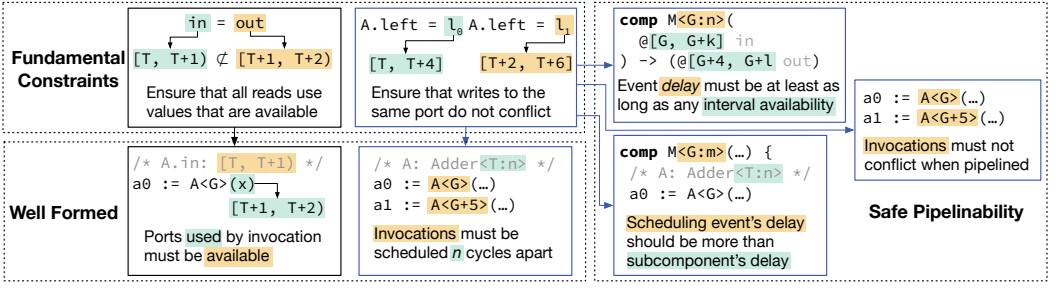

Fig. 5. Overview of the Filament type system. The fundamental constraints of hardware design imply other constraints. Well-formedness ensures that *one execution* of a component is correct. Safe pipelining ensures that *pipelined executions* of the component are correct.

***Ordering constraints.*** In order to capture the full expressivity of external components, Filament allows defining ordering constraints between events. For example, combinational components can provide a valid output for more than one cycle if the inputs are provided for multiple cycles. Therefore, a more precise interface of a combinational adder is:

```
comp Add<G: L-G, L: 1>(@[G, L] l: 32, @[G, L] r: 32) -> (@[G, L] o: 32) where L > G
```

The events $G$ and $L$ mark the start and end for the input and output availability intervals. In order to ensure that the interval $[G, L]$ is well-formed, the signature requires $L > G$. The component guarantees that the output is provided for as long as the inputs are provided.

***Parametric delays.*** The new signature of adder additionally specifies a *parametric delay* of $L - G$ cycles to signal that the adder may not be reused while it is processing a set of inputs. In order to generate *static pipelines* which have input-independent timing behavior, Filament requires all such expressions to evaluate to a constant value. Like the example, an invocation of `Add` must provide some binding of the form $G = T + i$ and $L = T + k$ such that $k > i$, ensuring that the delay for the corresponding invocation is a compile-time constant $k - i$ and the ordering constraint $L > G$ is satisfied.

```
A := new Add;
// delay = (G+3)-G = 3
a0 := A<G, G+3>(x, y);
```

The signature of registers in Filament allows them to provide the output for as long as needed, similar to an adder. However, because a register is a state element, it only requires its input for one cycle. Furthermore, the delay signals that the register can accept a new write during the last cycle when the output is available.

```
comp Register<G: L-(G+1), L: 1>(
  @interface[G] go: 1, @[G, G+1] in: 32) -> (@[G+1, L] out: 32) where L > G+1;
```

## 4 TYPE SYSTEM

Filament's type system enforces two fundamental restrictions of hardware design:

(1) All reads only use *semantically* valid values. A port or wire will always have a value on it. Filament's availability intervals mark when the values are semantically valid.

(2) Writes do not conflict. This is a corollary of the property that uses of a resource must not conflict because use of a resource is represented through a write.

Filament ensures these properties using two checking phases: well-formedness checking, which ensures that a *single execution* of a component is correct, and *safe pipelining*, which ensures that *pipelined executions* of a component are correct.

### 4.1   Delay Well-Formedness

The delay of an event encapsulates all possible conflicts between parts of the pipeline scheduled using it. Filament requires that the delay of an event is at least as long as each interval that mentions it which ensures that instance reuse does not create conflicts between its input and output ports.

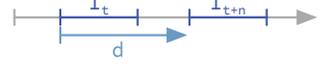

The proof is straightforward: for two invocations at time $t$ and $t + n$ such that $n \geq d$ where $d$ is the delay, let $I_t$ and $I_{t+n}$ be the availability intervals of the input $i$. Since we know that the start times of the intervals are at least $d$ cycles apart ($I_{t+n} - I_t \geq d$), and that length of the intervals is bounded by $d$ ($|I_t| \leq d$) we can conclude that they do not overlap.

### 4.2   Well-Formedness

***Valid reads***. In order to ensure this property, Filament needs to make sure that port values are only read when they are semantically valid. Signals are used in two places:

(1) *Connections* (Section 3.5) forward a value from one port to another. Filament ensures that the availability of the output port is at least as long as the requirement of the input port.
(2) *Invocations* (Section 3.4) schedule the use of a component instance using a set of events. Checking the validity of an invocation boils down to two steps: the requirements of the instance's input ports can be computed by binding the event variables in its signature to the invocation's event. Next, each argument essentially represents a connection between the instance's input and the argument and is checked using the criteria for connections.

***Conflict-free***. If an invocation schedules an instance with delay $d$ using the event $G$, the instance may not be reused between $[G, G+d]$. This both ensures that the there are no conflicts between input and output ports (Section 4.1) and that none of the subcomponents conflict. The latter property holds because safe pipelining constraints ensure that a valid delay can correctly encapsulate all possible conflicts between subcomponents (Section 4.4). In the example program, the two invocations of `M` overlap causing Filament to reject this program.

```
comp Mult<T:3>(...);
comp main<G:10>() {
  M := new Mult;
  // busy b/w [G, G+3]
  a0 := M<G>(a, b);
  // busy b/w [G+1, G+4]
  a1 := M<G+1>(a0.out, b);
```

### 4.3   Initiation Intervals

Pipelining is an important optimization since it allows a module to process multiple inputs in parallel. For example, a multiplier with a three cycle latency, but an *initiation interval* of one cycle takes three cycles to compute an output but can accept new inputs every cycle. In Filament, the delay of an event corresponds to initiation interval. While hardware designers talk about initiation intervals of a component, Filament generalizes it by allowing a component to have multiple events. In this case, each event specifies the initiation interval of some part of the internal pipeline. Filament ensures that the delay of a module describes a valid initiation interval, defined as follows:

*Definition 4.1 (Initiation Interval).* Let $P(t)$ be the execution of pipeline $P$ at time $t$. $P(t_0) \perp P(t_1)$ states that the pipeline executions of $P$ at $t_0$ and $t_1$ do not have resource conflicts. Then $I$ is a valid initiation interval of pipeline $P$ if and only if

$$\forall n \geq 0 \; P(t) \perp P(t + I + n)$$

This definition requires that the pipeline is able to accept new inputs after *any* amount of time after the initiation interval. There might be other delays smaller than the initiation interval which allow the pipeline to accept new inputs in a small window of time before becoming invalid again.

This would correspond to the following definition of an initiation interval $I$:

$$\forall k \neq 0 \; P(t) \perp P(t + k \times I)$$

Filament uses the first definition because delays are also used to check the well-formedness constraints of a component. If we used the second definition, the well-formedness constraint would require that if an instance is scheduled at time $t$, it may only be scheduled again at other times $k * t$ which we think is less compositional. Regardless, this is not a fundamental limitation since both definitions can be encoded and enforced.

### 4.4 Safe Pipelining

While well-formedness ensures that one execution of a module is correct, i.e., all reads use valid values and there are no conflicts, safe pipelining must ensure that *pipelined executions* of the component do not create any additional conflicts. Checking that pipelined executions do not conflict is very similar to checking that invocations of the same instance do not conflict. This is because pipelined execution is exactly the same—an instance being reused after a period of time. Filament must show that for an invocation scheduled using event $G$, another invocation scheduled at any time after $G + d$ (where $d$ is the delay) does not conflict with the first invocation. The following checks are sufficient to prove this.

***Triggering Subcomponents.*** Filament requires that when an event is used to invoke a subcomponent, the event's delay must be at least as long as the delay of the subcomponent's event.

```
comp Mult<G:3>(@interface[G] go: 1, ...)
comp main<T:1>(@interface[T] go: 1, ...) { M := new Mult; m0 := M<T+2>(...) }
```

The event $T + 2$ is used to schedule the invocation of instance $M$ which has a delay of 3. However, $T + 2$ has a delay of 1, same as $T$. This is problematic because `main` may trigger every cycle while `M` can only support computations every 3 cycles. Filament therefore rejects this program.

***Reusing Instances.*** Previous checks already ensure that: (1) shared invocations do not conflict during one execution of the pipeline, and (2) pipelined execution of an invocation does not conflict with itself. However, we also need to ensure that pipelined invocations of a shared instance do not conflict with each other.

```
comp Mult<G: 3>(...)
comp main<T: 3>(...) {
    M := new Mult;
    m0 := M<T+2>(...);
    m1 := M<T+10>(...);
```

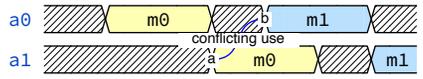

The example program will pass all our previous checks but is erroneous: executing the pipeline at time $T$ and $T + 1$ will cause the `m1` from the time $T$ execution to conflict with `m0` from the time $T + 10$ execution. Because Filament's definition of initiation interval allows re-execution at *any time* in the future, we must require that all invocations of a shared instance complete before the pipelined execution begins. The following is sufficient to ensure this: the delay must be greater than the number of cycles between the start of the earliest invocation and the end of the last invocation of a shared instance.

***Dynamic Reuse.*** Since Filament components can be parameterized by multiple events, it is possible to invoke an instance using two different events. In the example program, the type-checker would have to prove that the intervals $[G, G + 3)$ and $[L, L + 3)$ do

```
comp Dyn<G: ??, L: ??>(..) {
    M := new Mult;
    a0 := M<G>(a, b);
    a1 := M<L>(a0.out, b); }
```

not overlap to enforce conflict freedom. The constraint $L \geq G + 3$ is sufficient to prove this. However, there is no way to statically pipeline this module: the delay of $G$ is *dynamic*, it depends on exactly which cycle $L$ is provided which cannot be known a priori. There is no compile-time constant value that can express the delays for both events. This is because delays describe the timeline for a single event whereas dynamic modules require relating multiple events. Filament's solution is to disallow ordering constraints between events in user-level components which disallows the example program. External components (Section 3.6) can still use ordering constraints, but such constraints can only be satisfied using the natural order defined on $G + n$ events. This means in a well-typed program:

(1) All delays evaluate to compile-time constants.
(2) Invocations of a shared component all use the same event.

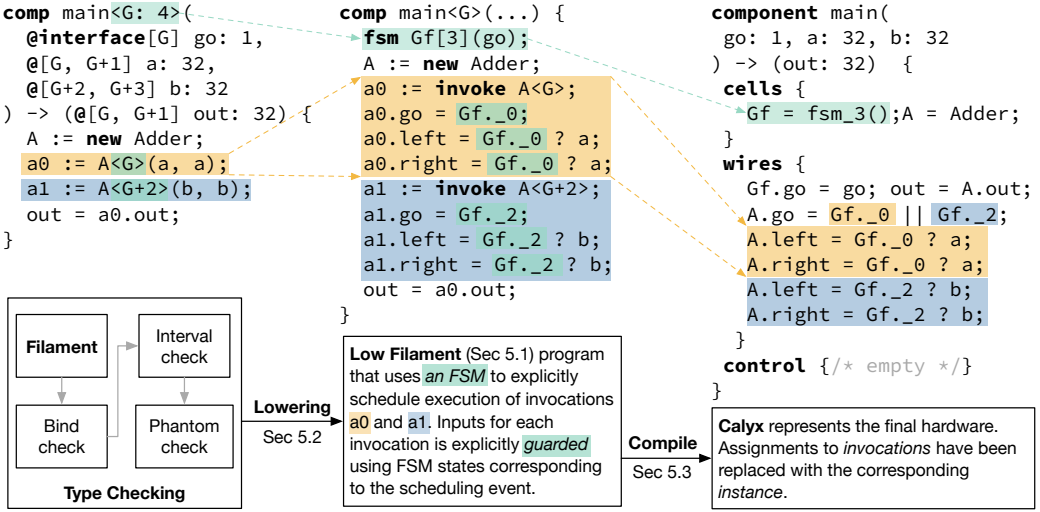

Fig. 6. Compilation Flow. Filament programs are type checked (Section 4) and lowered to *Low Filament* (Section 5.1) programs. Lowering (Section 5.2) instantiates explicit FSMs to schedule invocation. Finally, Low filament programs are compiled to Calyx [40] which optimizes the design and generates hardware circuits.

These constraints allow the compiler to generate efficient, statically timed pipelines from well-typed programs. Extending Filament with safe dynamic pipelines is an avenue for future work.

## 5 COMPILATION

Figure 6 shows an overview of the compilation flow. The primary goal of Filament's compilation pipeline is to transform the abstract schedules of invocations into explicit, pipelined control logic. The compiler first lowers programs into *Low Filament* which is an untyped extension of the Filament language that explicitly uses pipelined finite state machines (FSMs) to coordinate the execution of a module. Next, the compiler translates the program into the Calyx intermediate language [40] which performs generic optimizations and generates circuits.

### 5.1 Low Filament

Low Filament is an untyped version of Filament that introduces new constructs to explicitly represent the pipelined execution of a module.

***Explicit Invocations.*** Low Filament requires all ports corresponding to an invocation to be explicitly assigned. This includes interface ports, which high-level Filament manages implicitly.

**Guarded Assignment.** Filament uses *guarded assignments* to express multiplexing of signals and correspond directly to guarded assignments in Calyx [40]. The assignment only forwards the value from `out` when the guard is active. Otherwise, the value forwarded to `in` is undefined. Calyx's well-formedness condition requires that only one of the guards is active at a time for any given source port.

```
in = g1 ? out;
in = g2 ? out;
```

**Finite state machines.** Low Filament also provides the `fsm` construct to explicitly instantiate a pipeline FSM. This defines the FSM `F` with $n$ states and a single input port `trigger` which triggers its execution. This generates a shift-register of size $n$ with ports: `F._0, ..., F._{n-1}`. If *trigger* is set to 1 at event $G$, the port `F._i` will become active at event $G + i$.

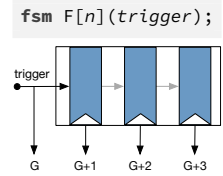

## 5.2 Generating Explicit Schedules

The compilation from Filament to Low Filament ensures that all high-level invocations have been compiled into explicit invocations. Figure 6 shows the compilation process for a program that uses an adder (`A`) through two invocations (`a0` and `a1`).

**FSM Generation.** The compiler instantiates an FSM for each event parameterizing the module. The example program uses event $G$ to schedule the invocations. The compiler walks over all expressions $G + i$ in the program to compute the number of stages for the pipelined FSM. While the original program does not explicitly mention the event $G + 3$, it is implied by the output port `a1.out` which is active in the interval $[G + 2, G + 3]$. The compiler instantiates the FSM `Gf` with 3 states triggered by the go signal. Note that the delay of the FSM *does not* affect the generation of the FSM.

**Triggering Interface ports.** The compiler then lowers the invocations by generating explicit assignments to the adder's interface port go. The first invocation, scheduled at $G$, uses the port `Gf._0` to trigger the invocation while the second invocation, scheduled at $G + 2$, uses the port `Gf._2`.

**Guard Synthesis.** In order to ensure that assignments from the two invocations to the data ports `left` and `right` do not conflict, the compiler synthesizes guards for the assignments. If the input port of an invocation require inputs during the interval $[G + s, G + e]$, the compiler generates the guard `Gf._s || ... || Gf._e` for the guard. Since the program is well-typed, the guard expressions for each invocation are guaranteed to not conflict (Section 4).

## 5.3 Lowering to Calyx

Low Filament is intentionally designed to be close to Calyx, so compilation is straightforward. For each FSM size $n$, we generate a Calyx component and instantiate it for the corresponding Filament component. The FSM is simply a sequence of registers connected together. Since assignments to all ports are explicit in Low Filament, we can simply compile the invocations by replacing them with the corresponding instance name. In the example program, assignments to both `a0.left` and `a1.left` are compiled to assignments to `A.left`. Since Filament guarantees that the generated guards are disjoint, we can be sure that Calyx will generate correct FSMs.

## 5.4 Optimizing Continuous Pipelines

Continuous pipelines do not make use of a signal to indicate when their inputs are valid and instead, they continuously process inputs. We can express such pipelines in Filament using *phantom events* (Section 3.6). Phantom events do not have a corresponding interface port and therefore cannot be used to trigger invocations. Filament ensures that a phantom event is used correctly through its phantom check analysis which ensures:

$x \in vars \quad t \in events \quad p, q \in ports$

$M ::=$

$\quad \textbf{def } C\langle t : n \rangle(p_1 : \pi_1, \ldots, p_j : \pi_j)\{c\}$

$c ::= c_1 \cdot c_2 \mid p_d = p_s \mid x := \textbf{new } C$

$\quad \mid x := \textbf{invoke } x\langle T \rangle(p_1, \ldots, p_j)$

$T ::= t \mid T + n \quad \pi ::= [T_1, T_2]$

$\tau ::= \forall\langle t : n \rangle(p_1 : \pi_1, \ldots, p_j : \pi_j)$

(a) Abstract syntax

$[\![c]\!] : \mathcal{L} \to \mathcal{L} \qquad \mathcal{L} : \mathcal{T} \to \mathcal{R} \times \mathcal{W}$

$[\![p_d = p_s]\!](L) = \text{map}(\lambda(R, W). \text{ if } p_s \in W$

$\quad \text{then } (R\{p_s/p_d\}, W) \text{ else } (R, W), L)$

$[\![c_1 \cdot c_2]\!](L) = [\![c_1]\!](L) \cup [\![c_2]\!](L)$

(b) Log-transformer semantics

$$\frac{\Delta, \Lambda_1, \Gamma \vdash c_1 \dashv \Lambda'_1, \Gamma_1 \qquad \Delta, \Lambda_2, \Gamma \vdash c_2 \dashv \Lambda'_2, \Gamma_2}{\Delta, \Lambda_1 * \Lambda_2, \Gamma \vdash c_1 \cdot c_2 \dashv \Lambda'_1 * \Lambda'_2, \Gamma_1 \cup \Gamma_2}$$

(c) Composition judgement

Fig. 7. Formal semantics of Filament where command is defined as a *log-transformer*. Typing judgements track the active timeline of an instance and ensure they are used in a disjoint manner.

*Definition 5.1 (Phantom Check).* A phantom event $G$ is used correctly if:

(1) It is not used to share any instances.
(2) It is only used to invoke subcomponents that use phantom events.

First, resource sharing is disallowed because any pipeline that shares an instance must use some signal to trigger an internal FSM and track which use of the instance is currently active. Second, a phantom event is only available at the type-level and cannot be reified since there is no interface port. Therefore, only components that use phantom events can be invoked with a phantom event.

Filament defines two state primitives: a *register* and a *delay* component.

```
comp Register<G: L-(G+1), L: 1>(
  @interface[G] en: 1, @[G, G+1] in: 32
) -> (@[G+1, L] out: 32) where L > G+1;
```

```
comp Delay<G: 1>(
  @[G, G+1] in: 32
) -> (@[G+1, G+2] out: 32);
```

As the type signatures denote, the difference is that a register can hold onto a value for an arbitrary amount of time while a delay can only hold onto a value for a single cycle. The `Delay` component accepts inputs every cycle and can therefore provide the output for one cycle. In contrast, the register can use the `en` signal to hold onto a value for an arbitrary amount of time.

***Compilation.*** The compiler does not instantiate FSMs or synthesize guards for invocations triggered using phantom events. Since Phantom Check ensures that all subcomponents themselves do not have an interface port, the compiler does not have to generate assignments for them. Filament generated code for continuous pipelines matches expert-written code.

## 6 FORMALIZATION

Figure 7a presents a simplified syntax for Filament: all components can be parameterized using exactly one constraint and cannot specify any ordering constraints between events. Since Filament disallows any form of event interaction in user-level components, multi-event user-level components are not fundamentally more expressive. Multi-event external components are more expressive but not supported in our formalism. A Filament program ($\mathcal{P}$) is a sequence of components which define a signature and a body in terms of commands: composition, connection, instantiation, and invocation.

## 6.1 Semantics

Figure 7b presents Filament's semantics which is defined as functions over logs ($\mathcal{L}$). A log maps events ($\mathcal{T}$) to a set of ports that are read from ($\mathcal{R}$) and a multiset of ports that are written to ($\mathcal{W}$). Intuitively, a log captures all the reads and writes performed during every cycle of a component's execution. We track the multiset of writes to capture conflicts—if there are multiple writes to the same port in the same cycle, then the program has a resource conflict.

Concrete logs are generated by the semantics of component definitions while commands simply transform them. For example, a port connection forwards the value from the source port $p_s$ to the destination port $p_d$. We model this by substituting all occurrences of $p_d$ to $p_s$ in the read set $\mathcal{R}$ when $p_s$ is defined in the write-set $\mathcal{W}$ and mapping it over all defined events in the log. Composition reflects the parallel nature of hardware—it simply unions the two logs together. Write conflicts can appear due to composition. The semantics of a program is the log generated by executing a distinguished main component with the empty log.[3] We formalize the well-formedness (Section 4.2) and safe pipelining (Section 4.4) constraints of the type system using this semantics.

*Definition 6.1 (Well-Formedness).* A component $M$ is well-formed if and only if its log is well-formed. A log $L$ is well-formed if and only if, for each event:
- There are no conflicting writes: $W_s = W$ where $W_s$ is the deduplicated set of writes.
- Reads are a subset of writes: $R \subseteq W_s$

*Definition 6.2 (Safe Pipelining).* If a component $M$ has an event $T$ with delay $d$, and $[\![M]\!]_G$ represents its log where $T$ is replaced with the event $G$, then $M$ is safely pipelined if and only if all logs $L_n$ are well-formed: $L_n = \forall n \geq d \;\; [\![M]\!]_T \cup [\![M]\!]_{T+n}$

## 6.2 Type System

Filament implements a type system inspired by separation logic [46] to enforce the well-formedness and safe pipelining constraints. Our presentation focuses on the specific typing judgement that ensures that there are no conflicting uses of an instance. Appendix A.3 provides the full type system. At a high level, our typing judgement for composition (Figure 7c) mirrors the parallel composition rule used in concurrent separation logic [9]—the two commands are checked under two disjoint resource contexts. Our insight is adapting the definition of separating split to timelines of instances and ensuring that instance reuse does not conflict. The typing judgements have the form: $\Delta; \Lambda; \Gamma \vdash c \dashv \Lambda'; \Gamma'$. $\Gamma$ is the standard type environment, $\Delta$ tracks each event's delay, and $\Lambda$ is the *resource context*.

***Resource contexts and separating split.*** $\Lambda$ is the resource context and tracks the availability of each instance and port in the form of an interval ($\pi$). After instantiation, each instance is available in the interval $[0, \infty)$. The invocation rule (not shown) checks that, for an instance's event with a delay $d$, the instance is available in the interval $[G, G + d]$ where $G$ is the scheduling event. The composition rule (Figure 7c) splits the resource context before checking the two commands:

$$\Lambda = \Lambda_1 * \Lambda_2 \text{ iff } \forall (x : \pi) \in \Lambda \Rightarrow \exists \pi_1, \pi_2. (x : \pi_1) \in \Lambda_1 \wedge (x : \pi_2) \in \Lambda_2 \wedge \pi_1 \cap \pi_2 = \emptyset \wedge \pi_1 \cup \pi_2 = \pi$$

A valid split is one where the resulting contexts have disjoint intervals for each instance and the union of the intervals is the original interval. By using this definition of split, Filament ensures that invocations reuse instances in a non-conflicting manner. Our accompanying technical report presents the remaining type judgements that encode constraints to enforce well-formedness and safe pipelining and proves the following type soundness theorem (Appendix A.4):

THEOREM 6.3. *If $\Delta; \Lambda; \Gamma \vdash c \dashv \Lambda'; \Gamma'$ then $[\![c]\!]$ is well-formed (Definition 6.1).*

---
[3]The full semantics is provided in Appendix A.

## 7 EVALUATION

We evaluate Filament's ability to efficiently express a number of accelerator designs and to express the interfaces generated by state-of-the-art accelerator generators. Our evaluation answers the following questions:

(1) Can Filament express the interfaces generated by state-of-the-art accelerator generators and integrate with existing tools?
(2) Can Filament be used to generate efficient accelerators?

***Implementation.*** The Filament compiler is implemented using a pass-based compiler in 5426 lines of Rust, 341 lines of Verilog for the standard library primitives, and the latest version of the Calyx compiler [40] to generate Verilog. All benchmarks compile in under a second.

### 7.1 Expressivity Evaluation

To demonstrate the expressivity of Filament, we focus on giving type signatures to designs generated by Aetherling [22].

***Aetherling's space-time types.*** Aetherling [22] is a functional, dataflow DSL that generates statically-scheduled, streaming accelerators for image processing tasks. Aetherling's "space-time" types enable users to express the shape of the data stream as a sequence of valid and invalid signals. For example, the type `TSeq 1 1` denotes that there will be a stream with one valid element followed by one invalid element. Nesting these types allows users to express more complex shapes: `TSeq 3 0 (TSeq 1 1)` denotes that there will be three valid elements, with no invalid values, each of which has a shape described by `TSeq 1 1`. In our case study, we import 14 designs implementing two kernels: conv2d and sharpen. Aetherling's evaluation studies 7 design points for each kernel with different resource-throughput trade-offs. Filament can express the interface types for all designs and, in the process, finds several bugs in the generated interfaces.

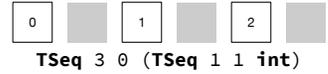

***Cycle accurate harness.*** We implemented a generic, *cycle-accurate* harness to test Filament programs. At a high-level it:

(1) Provides the inputs for exactly the cycles specified in a component's interface.
(2) Pipelines the execution of the component using event delays.
(3) Captures the value of output ports in the intervals provided in the signature.

The harness extracts the availability intervals and the event delays using a simple command-line flag provided to the compiler and executes the design using the cocotb Python library [17]. The design of this generic harness is reliant on a Filament-like system to document the timing behavior of modules; without Filament, a user would have to manually extract this information from the Verilog code.

***Methodology.*** We compile each Aetherling design to Verilog and use Aetherling's command line interface to extract the design's latency information. Each benchmark has five fully-utilized designs, which can accept new inputs every cycle, and two underutilized designs which produce 1/3 and 1/9 pixels per clock cycle and accept new inputs every 3 and 9 cycles. We give each design a type signature and validate its outputs. For designs with mismatched outputs, we change the latency till we get the right answer.

***Latency.*** Table 1 reports the latencies as provided by Aetherling's command line interface and those that we found to generate correct outputs with Filament's cycle accurate test harness. Of the 14 designs, Aetherling reports incorrect latencies for 5 designs.

| Throughput | Reported | Actual | | Throughput | Reported | Actual |
|:---:|:---:|:---:|---|:---:|:---:|:---:|
| 16 | 7 | 7 | | 16 | 7 | 7 |
| 8 | 6 | 6 | | 8 | 7 | 7 |
| 4 | 6 | 6 | | 4 | 7 | 7 |
| 2 | 6 | 6 | | 2 | 7 | 7 |
| 1 | 7 | 7 | | 1 | 8 | 8 |
| 1/3 | **10** | 12 | | 1/3 | **11** | 13 |
| 1/9 | **16** | 21 | | 1/9 | **17** | 20 |
| (a) Reported latencies for conv2d | | | | (b) Reported latencies for sharpen | | |

Table 1. Latencies of Aetherling Designs. Highlighted latencies are reported incorrectly by Aetherling.

***Underutilized designs.*** Aetherling explores the utility of *underutilized* designs which produce less than one pixel per clock cycle. Aetherling's compiler optimizes such designs by sharing compute resources. An Aetherling design that produces 1/9 pixels per clock has the type `TSeq 1 8 uint8` which states that there will be 1 valid datum followed by 8 invalid ones. The type indicates that the design generated by Aetherling should only use its input in the first cycle since the data provided in the next cycles is invalid. However, this interface is incorrect.

```
comp Conv2d<G: 9>(
  @[G, G+6] I: 8,
) -> (@[G+21, G+22] O: 8);
```

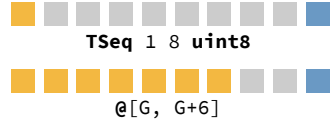

The Filament type, which reflects the actual interface needed to correctly execute the module, requires the design to hold its input signal for six cycles, i.e., the data element must be valid for six cycles instead of just one; the Aetherling implementation breaks its own interface. The Aetherling test harness does not catch this bug because it always asserts all inputs for 9 cycles. In contrast, Filament's test harness only asserts the input signal for as long as the corresponding availability interval specifies. Finally, the delay for the phantom event *G* encodes the fact the design can process a new input every 9 cycles. This illustrates the subtlety of specifying time-sensitive interfaces which accurately describe signal availability and pipelining.

***Other designs.*** We also import designs generated from PipelineC [30], an open-source high-level synthesis compiler that transforms a C-like language into Verilog (Appendix B.2). Providing type signatures for these was straightforward since PipelineC always fully pipelines designs and prints out the design's latency on the command line.

## 7.2 Accelerator Design with Filament

We study Filament's efficacy in generating efficient designs and reusing components generated from other languages by implementing a two-dimensional convolution in Filament. We build two Filament-based designs and compare them to the Aetherling-generated conv design.

***Architecture.*** Our implementation is directly inspired by the structure of the Aetherling implementation of conv2d that outputs 1 pixel per clock cycle. The design uses a $3 \times 3$ filter over a $4 \times 4$ matrix. The `Stencil` module (Figure 8a) implements a line buffer to save the last 11 values and outputs 9 values corresponding to the filter start index. The `Conv2d` kernel takes 9 values as inputs and produces an output corresponding to the result of the convolution.

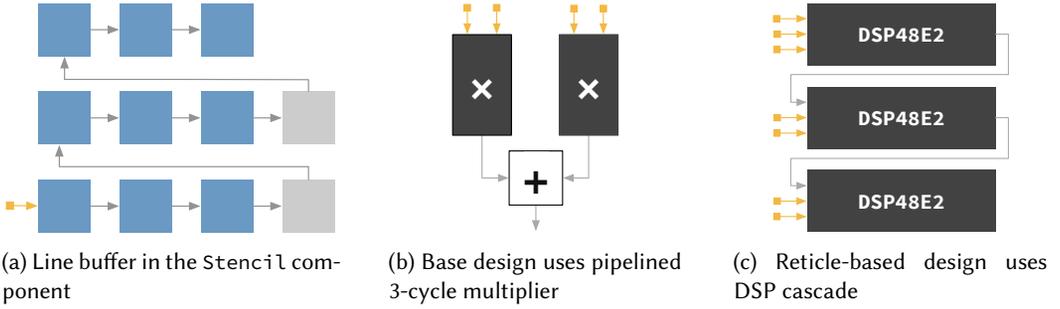

(a) Line buffer in the `Stencil` component

(b) Base design uses pipelined 3-cycle multiplier

(c) Reticle-based design uses DSP cascade

Fig. 8. Components used in the design of Filament-based conv2d convolution. The stencil component provides the last three inputs and is either connected to the naive multiplier or a Reticle-generated DSP cascade.

***Stream primitives in Filament.*** To implement line buffers, we implement a new `Prev` component which outputs the last value stored in it.[4] The Verilog implementation of `Prev` is simply a register but Filament gives it a different type signature—it allows access to the

```
comp Prev[SAFE]<G: 1>(
  @interface[G] en: 1,
  @[G, G+1] in: 32,
) -> (@[G, G+1] out: 32);
```

output in the same cycle when the input is provided which corresponds to reading the previous value in the register. The component uses a compile-time parameter `SAFE` to indicate whether the first read produces an undefined value. We also define a `ContPrev` component which is similar to a `Prev` component but uses a phantom event and can therefore be used in continuous pipelines (Section 5.4). The stencil component (Figure 8a) is implemented as a sequence of `Prev` components.

***Design 1: Pipelined multipliers.*** The base `Conv2D` kernel uses fully pipelined multipliers with a three cycle latency and combinational adders. The multipliers do not have any associated Verilog implementation—they are implemented using Xilinx's LogiCORE multiplier generator [2]. However, Filament makes it easy to interface with them by providing a type-safe extern wrapper (Section 3.6).

***Design 2: Integrating with Reticle.*** Our second design uses a dot-product unit generated using Reticle [49], a low-level language for programming FPGAs. Figure 8c shows the architecture Reticle generates to make use of *DSP cascading* which efficiently utilizes resources present on an FPGA. DSP cascading explicitly instantiates low-level FPGA primitives and connects them together to implement the computation: $y = c + \Sigma_{i=0}^{3} a_i \times b_i$. Unlike standard compilation flows which rely on the synthesis tool to infer DSP usage from *behavioral* descriptions, Reticle generates *structural* descriptions that predictably map

```
comp Tdot<G: 1>(
  clk: 1, reset: 1,
  @[G, G+1] a0: 8,
  @[G, G+1] b0: 8,
  @[G+1, G+2] a1: 8,
  @[G+1, G+2] b1: 8,
  @[G+2, G+3] a2: 8,
  @[G+2, G+3] b2: 8,
  @[G+2, G+3] c: 8,
) -> (@[G+5, G+6] y: 8)
```

onto DSPs. We provide a type signature for the Reticle design which indicates that the inputs must be provided in a staggered manner. Note that this is not implementation details leaking through—a DSP cascade that starts a new computation every cycle needs to either register all its inputs or provide them in a staggered manner.

***Evaluation methodology.*** We validate the correctness of all the designs using our timing-accurate test harness and compare the area and latency of the designs. For each design, we increase the target frequency till we reach worst negative slack of less than $0.1 ns$ and synthesize them using Vivado v2020.2. Each design has a throughput of 1 pixel per clock cycle.

---

[4]`prev` is a common operator in dataflow and functional reactive languages.

***Summary.*** Table 2 shows the results of the comparison: the Filament design can be synthesized at a higher frequency and uses fewer resources than the Aetherling design. This is because Filament can safely and directly use low-level implementation mod-ules which can be directly compiled into a safe and efficient design. In contrast, the Aetherling compiler has to generate extra logic when bridging the gap between its high-level language and low-level circuits. The Reticle-based design uses an order of magnitude fewer logic resources than the base Filament design or the Aetherling design. This is because unlike Aetherling, Reticle generates low-level *structural* Verilog which can predictably map onto DSP resources. This demonstrates the utility of Filament as both an integration and design language—designs in Filament can use low-level hardware modules safely and compose complex modules generated from other languages. It also reveals another use case for Filament: instead of directly generating Verilog, Aetherling-like languages can generate Filament programs and enable performance engineers to optimize the designs further and remove abstraction overheads.

Table 2. Resource usage and frequency of conv2d designs. Best values highlighted.

| Name | LUTs | DSPs | Registers | Freq. (MHz) |
|------|------|------|-----------|-------------|
| Aetherling | 104 | 10 | 78 | 769.2 |
| Filament | 128 | **9** | **11** | **833.3** |
| Filament Reticle | **14** | **9** | 20 | 645.1 |

***Other designs.*** Appendix B.1 details the other designs implemented in Filament: (1) floating-point computations, and (2) matrix-multiply systolic array [33].

## 8  RELATED WORK

***Dataflow languages.*** Reactive dataflow programming languages [8, 12, 24] provide stream operators scheduled using logical time steps. Software dataflow languages [45, 48] provide high-level, declarative operations that can target multiple backends like CPUs and GPUs. Compiling these languages to hardware requires complex transformations [5, 44]. Filament is lower-level and easier to compile since it directly reasons about hardware modules and is appropriate as a target language for hardware generators for these languages.

***Accelerator design languages.*** Accelerator design languages [1, 18, 22, 25, 26, 31] provide high-level abstractions to design hardware accelerators. Filament is a low-level HDL that provides a type system to directly interface with hardware modules and is appropriate both as an integration language and as a target language for compilers for ADLs.

***Embedded HDLs.*** Embedded HDLs [3, 4, 6, 16, 29, 34] use software host languages for metaprogramming. Most eHDLs simply use the host language's type system to ensure simple properties like port width match and signedness. Filament's type system focuses on expressing structural and temporal properties of the hardware itself. Rule-based HDLs [7, 41] use *guarded atomic actions* to provide transactional semantics for hardware specification. To preserve their high-level semantics, the compiler must generate complex scheduling logic that dynamically aborts conflicting rules. In contrast, Filament specifies program schedules using invocations which are checked at compile time and predictably map to efficient, pipelined hardware.

***Type systems for Hardware Design.*** Dahlia [38] is a C-like language that uses a substructural type system to ensure that high-level programs do not violate hardware constraints. Dahlia's affine reasoning can be encoded in Filament's type system. Kami [13] is a proof-assisted framework for designing hardware. Kami can be used to prove full functional correctness of hardware modules but requires users to write proofs while Filament's type system is focused on timing properties and is automatic. Filament, in contrast, focuses on latency-sensitive interfaces in presence of pipelining.

Pi-Ware [43] uses dependent types to ensure low-level circuit properties such as ensuring all ports are connected, and wire sorts [14] check whether module composition can create *combinational loops*, both of which are orthogonal to Filament's guarantees.

***Session types***. Cordial [21] is a type system inspired by *session types* to reason about latency-insensitive hardware protocols while Ghica [23] presents game-semantics-inspired type system to model interfaces; both tools do not reason about pipelining. Das et al. [20] extend session types with temporal modalities from linear temporal logic, enabling it to express time-synchronous properties. While their logic can be used to express Filament's well-formedness property, it is not obvious how to encode the safe pipelining constraints since they reason about all possible conflicting pipelined executions. Formally connecting Filament to the session types literature is an avenue for future work.

***Timing specifications***. IP-XACT [28] is an XML-based interface definition language that is used to package up coarse-grained, DMA-based interfaces. Filament, in contrast, specifies the low-level timing behavior of hardware modules. Synopsys Design Constraints (SDC) [47] are commonly used to configure the mapping of RTL programs to a physical implementation. While SDC can be used to specify physical timing constraints such as clock period and delays for combinational paths, these properties correspond to the physical implementation of a circuit instead of temporal behavior of a module. Furthermore, both solutions focus on specification alone. In contrast, Filament can both specify and check timing behaviors.

***Model checking***. Model checking [15] is a popular technique to verify hardware designs. SystemVerilog Assertions [50] provide a linear temporal logic (LTL) based specification language to verify properties of hardware designs [10, 35]. Such systems provide whole program guarantees and can prove more general timing properties than Filament. Filament focuses on providing compositional guarantees and interface specifications. Filament signatures can be potentially compiled to LTL specification and checked by the aforementioned tools.

## 9   FUTURE WORK

Filament represents a first step toward a new class of type systems that can reason about structural and temporal properties of hardware designs.

***Dynamic pipelines***. While static pipelines encompass a large and interesting set of hardware designs, dynamic pipelines are important for expressing complex designs. Filament's characterization of pipelining constraints (Section 4) provides a formal foundation to reason about dynamic pipelines. A possible solution is allowing modules to *generate* events in addition to consuming them; this would correspond to an existential quantification over the event type as opposed to the universal quantification in the current system. The challenge is showing that this extension produces well-formed dynamic pipelines.

***Polymorphic Filament***. HDLs allow programmers to implement modules with parametric latencies. For example, a shift register implementation can use a parameter to specify the number of stages which affects its latency. A polymorphic extension to Filament would allow programmers to write down *generative modules*, similar to embedded HDLs, and use the type system to guarantee that *any valid assignment* the parameters will generate a safe pipeline.

***Type-preserving compilation***. High-level synthesis, the process of compiling imperative languages like C to hardware, is notoriously buggy. While previous efforts have focused on building fully verified compilers [27, 32], they fail to capture common pipelining optimizations such as modulo scheduling [19]. We believe that Filament can be used as a typed intermediate language for building a type-preserving compiler [36] that rules out pipeline-related HLS compilation bugs.

## 10 CONCLUSION

Unlocking the true potential of reusable hardware requires detailed understanding of the structure and timing of the implementation. Filament exposes this knowledge through interfaces and enables users to reuse designs and fearlessly build high-performance hardware.

## ACKNOWLEDGMENTS

We thank Drew Zagieboylo, Ryan Doenges, Stephen Neuendorffer, Andrew Appel, Christopher Batten, and Zhiru Zhang for insightful conversations. David Durst's help made it possible to reproduce Aetherling's evaluation and compare it with Filament. Many thanks to our anonymous reviewers who provided valuable feedback and pointers to related work.

This work was supported in part by the Center for Applications Driving Architectures (ADA), one of six centers of JUMP, a Semiconductor Research Corporation program co-sponsored by DARPA. It was also supported by NSF awards #1845952, #2124045, and #1909073 and gifts from Google and SambaNova.

## ARTIFACT

The artifact for this paper has been archived and made publicly available [39] in the form of a virtual machine. In order to reproduce our results, we recommend using *v2020.2*. Owing to the nondeterministic nature of hardware synthesis, we do not expect that reproduced results will exactly match Table 2, but we expect that the high-level takeaway, that Filament designs take fewer resources and run faster, can be reproduced.

$$
\begin{aligned}
&[\![ c ]\!] &&: \mathcal{L}og \rightarrow \mathcal{L}og \qquad (\mathcal{L}og = \mathcal{T} \rightarrow \mathcal{R} \times \mathcal{W}) \\
&[\![ x := \mathbf{new}\ C ]\!] &&= \mathrm{id} \\
&[\![ p_d = p_s ]\!] &&= \lambda(R, W).\ \mathrm{if}\ p_s \in W\ \mathrm{then}\ (R\{p_s/p_d\}, W)\ \mathrm{else}\ (R, W) \\
&[\![ c_1 \cdot c_2 ]\!] &&= [\![ c_1 ]\!] \cup [\![ c_2 ]\!] \\
&[\![ x_1 := \mathbf{invoke}\ x_2 \langle T' \rangle (q_1, \cdots, q_m) ]\!] &&= [\![ \mathrm{connects}(x_1, [q_1, \ldots, q_m]) ]\!] \circ [\![ x_2 ]\!] \\[1em]
&[\![ M ]\!] &&: \mathcal{P} \\
&[\![ \mathbf{def}\ C \langle t : n \rangle (i_0 : \pi_0, \ldots, o_0 : \pi_i, \ldots)\{c\} ]\!] &&= \{\pi_0 \mapsto i_0, \ldots\} \times \{\pi_i \mapsto o_0, \ldots\}
\end{aligned}
$$

Fig. 9. Log-transformer semantics for Filament's core language. Each command produces a log ($\mathcal{L}$) which maps events ($\mathcal{T}$) to multisets of reads ($\mathcal{R}$) and writes ($\mathcal{W}$). Component definitions produce partial logs ($\mathcal{P}$) by mapping availabilities of inputs to reads and availabilities of outputs to writes.

## A  FULL FORMALISM

### A.1  Syntax

We used a simplified version of Filament for our formalization: all components can be parameterized using exactly one constraint there cannot specify any ordering constraints between events. Neither simplification loses generality because user-level components with multiple events cannot define any form of interaction between them—they are functionally equivalent to multiple components with disjoint events.

A Filament program is a sequence of components $M$ each of which encapsulates the structure and schedule of a pipeline. Commands $c$ include composition, port connections (Section 3.5), component instantiation (Section 3.3), and invocations (Section 3.4). Component are parameterized using exactly one event and invocations allow scheduling using one event.

$$
\begin{aligned}
&x, C \in vars \quad t \in events \\
&p, q \in ports \\
&M ::= \\
&\quad \mathbf{def}\ C \langle t : n \rangle (p_1 : \pi_1, \ldots, p_j : \pi_j)\{c\} \\
&\quad c ::= c_1 \cdot c_2 \mid p_d = p_s \mid x := \mathbf{new}\ C \\
&\quad\ \mid x := \mathbf{invoke}\ x \langle T \rangle (p_1, \ldots, p_j) \\
&\quad T ::= t \mid T + n \quad \pi ::= [T_1, T_2]
\end{aligned}
$$

### A.2  Semantics

The basis of our semantics is logs of reads and writes. Intuitively, every command generates a function from events to (multi)sets of reads and writes indicating which ports were read or written to at that particular event. More formally, every command is interpreted as a function over logs as presented in Figure 9 which provides a denotation of Filament programs as a log transformer ($\mathcal{L}$) of events to multisets of port reads ($\mathcal{R}$) and port writes ($\mathcal{W}$). Since components are allowed to use exactly one event, say $T$, the log maps events such as $T, T + 1$, etc. to reads and writes to ports defined by the subcomponents. Instantiation does not affect the logs, while connections rewrite the logs by adding the LHS port to the set of writes for event where the RHS port is defined. Parallel composition is interpreted as the union of the logs produced by the two commands.

Finally, invocations produce a log-transformer derived from its signature: the input ports are added to the reads ($\mathcal{R}$) and the output ports to the writes ($\mathcal{W}$) for each event contained in their corresponding availability interval. Note, however, that the semantics given by the signature uses the "incorrect" ports, since it is using the ports given by the signature. We work around by using the connects metafunction which is defined was

$$\text{connects}(x_1, [q_1, \cdots q_m]) = (x_1.p_1 = q_1); \cdots ; (x_1.p_m = q_m),$$

where we are implicitly assuming that the ports $x_1.p_i$ have not been used. Intuitively, this metaprogram coverts the arguments to an invocation into connections and generates a log by interpreting them using the denotation over commands. The semantics of a program is defined by the log produced by a distinguished top-level component.

***Components and its semantics.*** For example, a combinational adder and a sequential multiplier with a two-cycle latency produce the following logs:

$$[\![\,\textbf{def } \text{add}\langle G : 1\rangle(\text{l}: [G, G+1], \text{r}: [G, G+1], \text{out}: [G, G+1])\,]\!] = G \rightarrow (\{\,\text{l}, \text{r}\,\}, \{\,\text{go}, \text{out}\,\})$$

$$[\![\,\textbf{def } \text{mul}\langle G : 2\rangle(\text{l}: [G, G+1], \text{r}: [G, G+1], \text{out}: [G+2, G+3])\,]\!] = G \rightarrow (\{\,\text{l}, \text{r}\,\}, \{\,\text{go}\,\})$$

$$G+1 \rightarrow (\emptyset, \{\,\text{go}\,\})$$

$$G+2 \rightarrow (\emptyset, \{\,\text{out}\,\})$$

Note that the use of the instances is reflected through the writes their interface ports go (not shown in the signature). The log indicates that the multiplier accepts new values every 2 cycles by writing to the go port in both cycles $G$ and $G+1$. Because we track multisets of reads and writes, we can track conflicting writes to the same port.

Using these semantics, we can define the well-formedness constraint (Section 4.2) on logs:

*Definition A.1 (Well-Formedness).* A log $\mathcal{L}$ is well-formed if and only if for all events

- There are no conflicting writes: $\mathcal{W}_s = \mathcal{W}$ where $\mathcal{W}_s$ is the deduplicated set of writes.
- Reads are a subset of writes for every event: $\mathcal{R} \subseteq \mathcal{W}_s$.

While in real hardware, values are always available on a port or a wire, Filament's semantics only track semantically valid values from a read. Usage of hardware resources is denoted by a write, and it is physically impossible to write two values to a port; instead, the circuit uses a multiplexer to select between the two values. Multiple uses of a resource silently corrupt the data.

The safe pipelining constraints (Section 4.4) can be defined in terms of repeated execution of the semantics of a program:

*Definition A.2 (Safe Pipelining).* If a component $M$ has an event $T$ with delay $d$, and $[\![M]\!]_G$ represents its log where $T$ is replaced with the event $G$, then $M$ is safely pipelined if and only if for every $n \geq d$ the logs $L_n = [\![M]\!]_T \cup [\![M]\!]_{T+n}$ are well-formed.

## A.3  Type System

Our presentation focuses on Filament's substructural type system that is used to track non-conflicting use resources as well as signal validity. We elide the description of features that track things such as port widths which are standard.

***Typing contexts.*** The typing judgements use the following typing contexts:

- $\Gamma$ tracks the types for components and instances and availability of ports.
- $\Delta$ tracks the delays associated with each event in the context.
- $\Lambda$ is the *timeline context* and tracks the availability of each instance and port.

$$\tau ::= \forall \langle t : n\rangle (p_1 : \pi_1, \ldots, p_j : \pi_j)$$
$$\Gamma ::= \cdot \mid \Gamma, C : \tau \mid \Gamma, p : \pi$$
$$\Delta ::= \cdot \mid \Delta, t : n$$
$$\Lambda ::= \cdot \mid \Lambda, I : \pi \mid \Lambda, p : \pi$$

The type context ($\Gamma$) and timeline context ($\Lambda$) store timelines for ports and instances respectively. Timelines for ports are *reusable* since reading a port does not consume it during that cycle. However, the timeline of an instance is *consumed* when it is used in a cycle. Because of this, timeline contexts also provide a *separating union* inspired by separation logic [46].

***Splitting timelines.*** A valid separating split of a timeline context $\Lambda = \Lambda_1 * \Lambda_2$ if and only if both $\Lambda_1$ and $\Lambda_2$ bind all the same instances and for each instance, the timelines are disjoint. Formally:

$$\Lambda = \Lambda_1 * \Lambda_2 \text{ iff } \forall (I : \pi) \in \Lambda \implies \exists \pi_1, \pi_2. (I : \pi_1) \in \Lambda_1 \land (I : \pi_2) \in \Lambda_2 \land \pi_1 \cap \pi_2 = \emptyset \land \pi_1 \cup \pi_2 = \pi$$

***Instantiating components.*** Instantiating a module binds the signature of the component to the instance and make the resource available throughout the timeline of the program, denoted by $[0, \infty)$.

$$\frac{\Gamma(C) = \tau \qquad \Gamma' = \Gamma, I : \tau \qquad \Lambda' = \Lambda, I : [0, \infty)}{\Delta, \Lambda, \Gamma \vdash I := \mathbf{new}\ C \dashv \Lambda', \Gamma'}$$

***Port connections.*** Connecting ports checks that the source port is available for at least as long as the destination port requires:

$$\frac{\Lambda(p_d) \subseteq \Gamma(p_s)}{\Delta, \Lambda, \Gamma \vdash p_d = p_s \dashv \Lambda, \Gamma}$$

***Splitting timelines with composition.*** The composition rule splits the timeline context using the separating split operator and checks the two commands. Note that that same type context $\Gamma$ is used for both commands which means previously defined ports are available in both the commands:

$$\frac{\Delta, \Lambda_1, \Gamma \vdash c_1 \dashv \Lambda_1', \Gamma_1 \qquad \Delta, \Lambda_2, \Gamma \vdash c_2 \dashv \Lambda_2', \Gamma_2}{\Delta, \Lambda_1 * \Lambda_2, \Gamma \vdash c_1 \cdot c_2 \dashv \Lambda_1' * \Lambda_2', \Gamma_1 \cup \Gamma_2}$$

***Checking invocations.*** The invocation rule enforces well-formedness and safe-pipelining constraints and is therefore quite verbose. We separate out type checking of invocations into three sets of premises that logically reflect the properties presented in Section 4.

$$\frac{\text{valid reads} \qquad \text{no conflicts} \qquad \text{safe pipelining}}{\Delta, \Lambda, \Gamma \vdash x := \mathbf{invoke}\, I \langle T \rangle (q_1, .., q_j) \dashv \Lambda, \Gamma''}$$

The first set of premises check that all the reads from all ports mentioned in an invocation are valid, i.e., they are available for at least as long as the instance's signature requires. Finally, invocations bind the availability of all the ports associated with the instance to the type context.

$$\Gamma(I) = \forall \langle t : n \rangle (p_1 : \pi_1, .., p_j : \pi_j) \qquad \Gamma(q_1) = \pi_1, .., \Gamma(q_j) = \pi_j$$
$$\pi_1' = \pi_1[t/T], .., \pi_j' = \pi_j[t/T] \qquad \Gamma(p_1) \subseteq \pi_1', .., \Gamma(p_j) \subseteq \pi_j'$$
$$\Gamma' = x.\{p_1 : \pi_1', .., p_j : \pi_j'\} \qquad \Gamma'' = \Gamma \cup \Gamma'$$

The next set of premises ensure that the instance is available in the current timeline context. This ensures that there are no conflicting uses of the component anywhere else in the design.

$$\Gamma(I) = \forall \langle t : n \rangle (p_1 : \pi_1, .., p_j : \pi_j)$$
$$\Lambda(I) = \pi \qquad [T, T + n] \subseteq \pi$$

The composition rule is responsible for selecting a valid split to ensure that the above rule's constraints are satisfied. If there is no such split possible, then the program has conflicting uses of the instance.

A final set checks for the safety of pipelining an invocation (Section 4.4):

$$\Gamma(I) = \forall \langle t : n \rangle (p_1 : \pi_1, .., p_j : \pi_j)$$
$$\mathcal{E}(T) = t' \qquad \Delta(t') \geq \Delta(t)$$

### A.4   Type Soundness

Our type system guarantees theorem focuses on the well-formedness property (Section 4.2). It states that well-typed commands preserve well-formed logs. A second soundness property of our semantics is that the log transformers generated by well-typed programs may only add available ports to the writes of logs. This is captured by the following theorem.

LEMMA A.3 (AVAILABILITY SOUNDNESS). *If $\Delta; \Lambda; \Gamma \vdash c \dashv \Lambda'; \Gamma'$, then for every log $\mathcal{L}$ an every event $T$, let $(\mathcal{R}, \mathcal{W}) = [\![c]\!] (\mathcal{L}, T)$ then $p \in \mathcal{W}$ if, and only if, $p \in \pi_2(\mathcal{L}(T))$ or $T \in \Lambda(p)$.*

PROOF. The proof follows by induction on the typing derivation. The first case is trivial, since the identity function maps any log to itself. The port connection case does not modify the write component, which makes it similar to the identity case. The composition operation follows by the inductive hypothesis. For the invocation case consider a port $p$ in the writes of the transformed log and assume that $p \notin \mathcal{L}$. By construction, this $p$ has to be one of the output ports which, by the typing rule, has to be available. The other direction follows by case analysis.                        □

By specializing the theorem above to the composition $c_1 \cdot c_2$ and by using the fact that, by assumption, the $\Lambda$ contexts of $c_1$ and $c_2$ are disjoint it follows:

COROLLARY A.4 (DISJOINT WRITES). *If $\Delta; \Lambda; \Gamma \vdash c_1 \cdot c_2 \dashv \Lambda'; \Gamma'$, then for logs $(\mathcal{R}_1, \mathcal{W}_1) = [\![c_1]\!] (\mathcal{L})$ and $(\mathcal{R}_2, \mathcal{W}_2) = [\![c_2]\!] (\mathcal{L})$, $\mathcal{W}_1 - \mathcal{L}$ and $\mathcal{W}_2 - \mathcal{L}$ are disjoint.*

THEOREM A.5 (SOUNDNESS PROPERTY). *If $\Delta; \Lambda; \Gamma \vdash c \dashv \Lambda'; \Gamma'$ then if $\mathcal{L}$ is well-formed, then $[\![c]\!] - \mathcal{L}$ is well-formed as well.*

PROOF. The proof follows by induction on the typing derivation of $c$.

- Case INSTANTIATE: by assumption, $\mathcal{L}$ is well-formed.
- Case CONNECTION: by assumption $\mathcal{L}$ is well-formed, therefore $\mathcal{W}$ is a set. For the first condition, there are two possibilities: either $p_d \in \mathcal{R}$ or $p_d \notin \mathcal{R}$. If the second case holds then, $\mathcal{R}\{p_d \mapsto p_s\} = \mathcal{R} \subseteq \mathcal{W}$. If $p_d \in \mathcal{R}$ then, by assumption of the typing rule, the availability of $p_d$ is a subset of the availability of $p_s$ and, by well-formedness of the input log, $p_s \in \mathcal{W}$, which implies $\mathcal{R}\{p_d \mapsto p_s\} \subseteq \mathcal{W}$.
- Case COMP: by the induction hypothesis $\mathcal{L}_1 = [\![c_1]\!] (\mathcal{L})$ and $\mathcal{L}_2 = [\![c_2]\!] (\mathcal{L})$ are well-formed, which implies that $\mathcal{R}_1 \subseteq \mathcal{W}_1$ and $\mathcal{R}_2 \subseteq \mathcal{W}_2$. By monotonicity of the union operation, $\mathcal{R} = \mathcal{R}_1 \cup \mathcal{R}_2 \subseteq \mathcal{W}_1 \cup \mathcal{W}_2 = \mathcal{W}$ and the first well-formedness condition holds. Additionally, by corollary A.4 the writes for the two logs are disjoint, making $[\![c_1 \cdot c_2]\!] (\mathcal{L})$ well-formed.
- Case INVOKE: The only new writes done by the invocation rule are the output ports of the instance. By construction, they are a set. To prove that the reads are a subset of the writes, observe that the log generated by an instance is, on purpose, ill-formed because it uses placeholder names for the reads as writes. It is the job of the connects to ensure that the log will be well-formed. Note that, by definition, the semantics of port connection only alter the log if the source is in the writing log. Therefore, if we want to show that the reads are a subset of the writes, every guard in the semantics of connects must be true so that there are no placeholder ports in the log. This follows from the fact that, by construction, every port connection will be well-typed and, by Lemma A.3, we can conclude.

                                                                                                                □

# B  EVALUATION

## B.1  Filament Designs

We implemented several designs in Filament to evaluate its expressivitiy. Each design has a corresponding baseline that we extracted from either a handwritten implementation or a generated design.

***Floating-point add.*** We extracted a floating point adder implemented in Verilog.[5] The implementation provides both a single-cycle and a pipelined version. We then translated the pipelined implementation into a Filament design and used a fuzzing harness to ensure that the outputs of the implementation matched to the source. In the process of translation, we found several bugs in the original implementation. Each bug was caused by the pipelined design accessing signals from a previous pipeline stage. Such bugs were immediately obvious in Filament since the type checker disallows using signals from a previous stage to compute the value of a signal in the current stage. For example, in the second stage of the pipeline, the adder attempts to use a value from the previous stage:

```
// The suffix *_s2 indicates that the signal is used in the second stage.
if (exponent != 0) begin
  L1_s2 = {1'b1, Large_s2[22:1]};
end else begin
  L1_s2 = Large_s2;
end
```

In Filament, this bug is caught by the type checker:

```
M := new Mux[23];
l1_mant := M<G+1>(e2_is_zero_s2.out, large_mant_s2.out, normed_large.out);
                  Available in [G, G+1] but required in [G+1, G+2]
```

Using cycle-accurate harness, coupled with a simple fuzzer made it easy to find such bugs by differential testing of the combinational, pipelined, and Filament implementations. Our takeaway from the study was that *once a Filament design functionally agrees with a pipelined design, any changes in the Filament code to pipeline the design are unlikely to introduce new bugs due to the type checking.*

***Systolic Array.*** Systolic arrays [33] are a common design pattern for implementing linear algebra operations. They mimic the two-dimensional structure of the data by arranging the computation in a two-dimensional array of processing elements and exploit the data reuse between computations. We took the baseline systolic array reported in the Calyx [40] paper and implemented it in Filament. We used the `Prev` component implemented in Section 7.2 to express the computation of the systolic array.

```
comp main<G: 1>(
  @interface[G] go: 1, @[G, G+1] l0: 32, @[G, G+1] l1: 32, @[G, G+1] t0: 32, @[G, G+1] t1: 32,
) -> (
  @[G, G+1] out00: 32, @[G, G+1] out01: 32, @[G, G+1] out10: 32, @[G, G+1] out11: 32,
) {
  // Systolic registers that go from left to right
  r00_01 := new Prev[32, 1]<G>(l0); r00_10 := new Prev[32, 1]<G>(t0);
  r10_11 := new Prev[32, 1]<G>(l1); r01_11 := new Prev[32, 1]<G>(t1);

  // Connection registers to processing elements
  pe00 := new Process<G>(l0, t0);
```

---

[5]Source: https://github.com/suhasr1991/5-Stage-Pipelined-IEEE-Single-Precision-Floating-Point-Adder-Design.

```
    pe01 := new Process<G>(r00_01.prev, t1);
    pe10 := new Process<G>(l1, r00_10.prev);
    pe11 := new Process<G>(r10_11.prev, r01_11.prev);

    out00 = pe00.out; out01 = pe01.out; out10 = pe10.out; out11 = pe11.out;
}
```

The processing element `Process` performs a multiply-accumulate operation:

```
comp Process<G: 1>(
  @interface[G] go: 1, @[G, G+1] left: 32, @[G, G+1] right: 32) -> (@[G, G+1] out: 32) {
  // If acc does not contain a valid value, use 0
  acc := new Prev[32, 0]<G>(add.out);
  go_prev := new Prev[1, 1]<G>(go);
  mux := new Mux[32]<G>(go_prev.prev, acc.prev, 0);

  mul := new MultComb[32]<G>(left, right);
  add := new Add[32]<G>(mux.out, mul.out);

  out = add.out;
}
```

This process element will run at a lower frequency because it uses a combinational multiply (`MultComb`) instead of a pipelined multiply (`Mult`). However, it is straightforward to change this by using a pipelined multiply instead since it only changes the latency of the process module. Our alternate design uses a pipelined multiplier (`FastMult`) which changes the latency of the design.

```
comp Process<G: 1>(
  @interface[G] go: 1, @[G, G+1] left: 32, @[G, G+1] right: 32) -> (@[G+3, G+4] out: 32) {
...
  mul := new FastMult<G>(left, right);
}
```

## B.2 Type Signatures

We import designs from PipelineC [30] and give them type signatures in Filament.

***Floating-point add.*** The floating-point add generated by PipelineC is automatically pipelined by the compiler based on the frequency target provided by the user. We used the floating-point unit implemented in Appendix B.1 to differentially test the module and ensure that we provided the right type signature for the module.

```
comp FpAdd<G: 1>(
  clk: 1, @[G, G+1] my_pipeline_x: 32, @[G, G+1] my_pipeline_y: 32
) -> (
  @[G+6, G+7] my_pipeline_return_output: 32
);
```

***AES.*** We give the type signature for an AES module implemented in PipelineC and automatically pipelined by the compiler:

```
comp AES<G: 1>(
  clk: 1, @[G, G+1] state_words: 128, @[G, G+1] keys: 1280) -> (@[G+18, G+19] out_words: 128)
```